# Sequence-Based Deep Learning for Handover Optimization in Dense Urban Cellular Network

Muhammad Kabeer[1,2], Rosdiadee Nordin[1,3], Mehran Behjati[1], Lau Sian Lun[1]

1. School of Engineering and Technology, Sunway University, No. 5, Jalan Universiti, Bandar Sunway 47500, Selangor, Malaysia

2. Department of Computer Science, Federal University Dutsinma, Katsina Nigeria

3. Future Cities Research Institute, Sunway University, No. 5, Jalan Universiti, Bandar Sunway 47500, Selangor, Malaysia

**Abstract**

Efficient handover management remains a critical challenge in dense urban cellular networks, where high cell density, user mobility, and diverse service demands increase the likelihood of unnecessary handovers and ping-pong effects. This paper leverages a real-world, multi-operator drive-test dataset of 30,925 labelled records collected within a 2 km² area around Sunway City to investigate sequence-based deep learning approaches for handover detection and avoidance. We formulate handover prediction as a sequence problem and evaluate Gated Recurrent Unit (GRU), Long Short-Term Memory (LSTM), and Transformer architectures under Reference Signal Received Power (RSRP)-only and all-feature settings. The integration of multi-dimensional features significantly enhanced handover performance in dense urban cellular networks. The proposed GRU-based model achieved a remarkable 98% reduction in ping-pong handovers, alongside a 46.25% decrease in unnecessary handovers, outperforming the baseline RSRP-only approach which yielded a 22.19% reduction. Furthermore, the model demonstrated a 46% improvement in Time of Stay (ToS), indicating more stable user connections. With an inference time of just 0.91 seconds, the solution proves highly efficient and well-suited for real-time edge deployment scenarios. Compared to the conventional 3GPP A3 algorithm, these improvements demonstrate significant gains in mobility robustness and user Quality of Experience (QoE) improvement. The dataset is released to foster reproducibility and further research in intelligent mobility management for 5G and beyond.

## 1. Introduction

The evolution of cellular networks from LTE to 5G has enabled unprecedented levels of connectivity, supporting applications such as real-time video streaming, cloud gaming, and intelligent transportation systems. However, the rapid densification of networks in urban areas has significantly increased the complexity of mobility management. In dense deployments, overlapping small cells and heterogeneous frequency layers create volatile radio conditions, particularly in environments with high vehicular speeds, complex pedestrian pathways, and irregular building topologies [1], [2].

A central challenge in this context is handover management. The handover process ensures that user equipment (UE) maintains seamless connectivity when moving between cells. Traditional methods, such as the 3GPP Event A3 mechanism, rely on threshold-based triggers and hysteresis margins. While simple and widely deployed, these rules are reactive and limited in adapting to rapidly changing urban environments. As a result, networks frequently experience unnecessary handovers, where transitions occur but do not improve service quality, and ping-pong effects, where UEs bounce back and forth between adjacent cells within short intervals. Both issues increase signaling overhead, drain device energy, and reduce QoE [3], [4].

Recent research has sought to overcome these limitations using data-driven methods. Early work applied machine learning techniques such as random forests and Gaussian processes for coverage and handover prediction [5]. However, these approaches generally treat handovers as independent events, neglecting the sequential dependencies inherent in user mobility and temporal radio dynamics. More recently, sequence-based deep learning models, such as LSTM, GRU, and Transformers, have emerged as powerful tools for modelling temporal dependencies in time-series data. These models have shown strong performance in mobility prediction and service migration tasks [6], [7], [8]. Hence, there is need to evaluate such models using realistic, multi-operator drive-test datasets in handover management.

This paper addresses these gaps by presenting a new urban, multi-operator drive-test dataset collected around Sunway City, Malaysia, comprising 30,925 labelled records and 132 empirically cellular nodes. The dataset includes serving and neighboring cell metrics Reference Signal Received Power (RSRP), Reference Signal Received Quality (RSRQ), Signal-to-Noise Ratio (SNR) etc, spatiotemporal features, and traffic-aware sessions (FTP, 1080p video streaming, HTTP). Leveraging this dataset, we evaluate three state-of-the-art sequence models GRU, LSTM, and Transformer for handover detection and avoidance. The models are assessed not only by standard classification metrics accuracy, precision, recall, F1-score but also by their operational impact on reducing unnecessary handovers, mitigating ping-pong events, and improving energy efficiency. The contributions of this paper are threefold:
1. A publicly available, real-world multi-operator dataset collected in Sunway City, enriched with spatiotemporal and traffic-aware features relevant to handover management.
2. A comparative study of GRU, LSTM, and Transformer models for handover detection and avoidance, formulated as a sequence prediction problem.
3. Empirical evaluation of the models in terms of ping-pong reduction, unnecessary handover reduction, and improvement in Time of Stay (ToS), providing practical insights for deployment in dense urban cellular networks.

## 2. Related Work

### 2.1 Handover Management in Cellular Networks

Mobility management has long been a cornerstone of cellular networks, ensuring seamless connectivity as users traverse multiple cells. The most widely deployed mechanism is the 3GPP Event A3 handover shown in *Figure 1*, where a handover is triggered when the RSRP of a neighboring cell exceeds that of the serving cell by a hysteresis margin for a predefined Time-to-Trigger (TTT). While effective in sparse deployments, the A3 mechanism struggles in dense urban environments. Heterogeneous fluctuations in signal quality caused by multipath fading, building shadowing, and heterogeneous small cell layouts lead to frequent unnecessary handovers and ping-pong effects, which degrade QoE and increase energy consumption due to excessive signaling [2], [3].

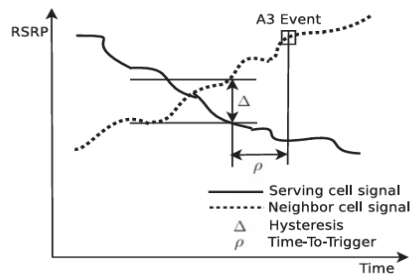

Figure 1 3GPP Handover Mechanism

### 2.2 Datasets for Cellular Network Analysis

Reliable datasets are essential to evaluate and improve handover strategies. Early studies focused on rural or suburban drive-tests, limiting their applicability to dense urban contexts. For instance, [9] released a 5G dataset capturing throughput and contextual metrics, while [10] presented a drive-test evaluation of mobile broadband in Malaysia. More recently, [11] introduced a multi-device, multi-operator dataset to capture coverage patterns in the Amazon region. However, these datasets often lack neighboring-cell measurements, traffic-aware profiles, and mobility labels, features necessary to realistically model and evaluate handover performance in complex urban settings.

### 2.3 Machine Learning in Mobility Prediction

To overcome the limitations of traditional threshold-based approaches, machine learning (ML) has been employed for coverage and handover prediction. For example, [5] applied supervised ML algorithms such as Random Forests for coverage estimation, while [12] explored UAV-assisted gap detection using ML. These methods show promise but often ignore sequential dependencies inherent in user mobility. Deep learning, particularly sequence-based architecture, has advanced the state of the art by capturing temporal dependencies in mobility data. LSTM and GRU networks have been applied to predict user trajectories and anticipate handover events, showing improved accuracy compared to static ML models [6], [8]. Recently, Transformers have demonstrated superior scalability and long-range modelling ability, making them suitable for trajectory prediction and mobility-aware handover [7]. Reinforcement learning approaches, such as Double Deep Q-Learning, further optimize handover policies by directly minimizing ping-pong and unnecessary handovers [4]. However, comparative evaluations of GRU, LSTM, and

Transformer under real-world, multi-operator drive-test conditions remain limited, leaving a gap that this study addresses.

## 3. Methodology

This study employs a three-stage methodology that integrates dataset collection, algorithm design for unnecessary and ping-pong handover detection and avoidance, and advanced deep learning model training and analysis as depicted in *Figure 2*. The workflow is designed to ensure reproducibility, robustness, and practical deployment in dense urban cellular environments.

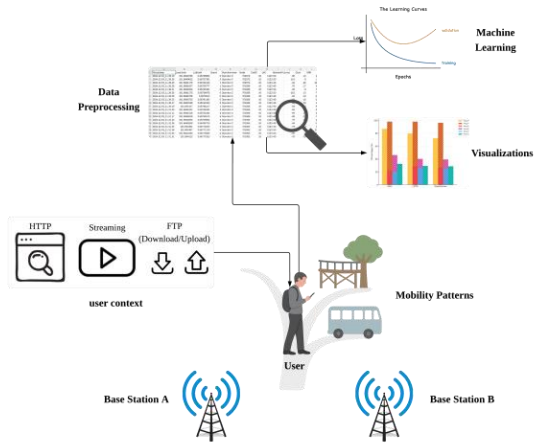

Figure 2 Handover Management Methodology

### 3.1 Dataset Collection

A real-world multi-operator dataset was collected within a 2 km² radius around Sunway City, Malaysia, covering heterogeneous mobility scenarios including pedestrian canopy walks, shuttle buses, and the Bus Rapid Transit (BRT) system, the coverage area is shown in *Figure 3,* which also serve as a heatmap showing RSRP distribution. Data collection was conducted for 3 major network operators which were anonymized as Operator A, B and C using calibrated Samsung S21 Ultra smartphones, selected for their ability to capture all 5G variants. Network testing applications Nemo Handy and GNetTrack Pro were benchmarked, with the latter chosen for its superior logging of handover-relevant features.

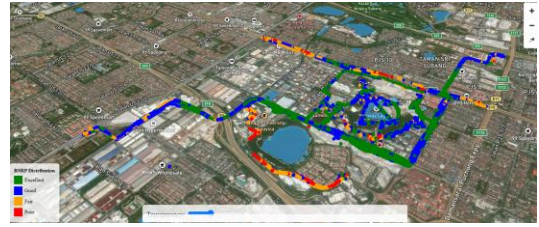

Figure 3 Dataset Collection Sites

The dataset comprises 30,925 labelled records [13] spanning three major Malaysian operators. Each record integrates Radio features: Serving and neighboring cell RSRP, RSRQ, and SNR. Spatiotemporal features: GPS coordinates, velocity, and bearing. User activity sessions: FTP, 1080p video streaming, and HTTP browsing, and Mobility contexts: Walking, shuttle, and BRT commute.

To ensure reliability, measurements were validated against 3GPP-defined thresholds. Missing values were handled through a structured pipeline, linear interpolation, and one-hot encoding or embeddings for categorical variables.

### 3.2 Detection and Avoidance Algorithm

The proposed framework formulates handover stability assessment as a sequence prediction problem, incorporating radio signal dynamics, trajectory information, and predicted ToS. Two complementary algorithms; Detection and Avoidance were designed, as summarized in Algorithms 1 and 2.

#### 3.2.1 Detection of Unnecessary and Ping-Pong Handovers

Given an input sequence of radio features (RSRP, SNR, slopes) and predicted ToS, the detection algorithm identifies unstable handovers characterized by short ToS and signal oscillations. A class-weight adjustment is applied to compensate for the imbalance between frequent stable handovers and rare ping-pong events.

**Algorithm 1: Detection**
Input: *Seq X (RSRP, SNR, Slopes), ToS y_p, PP Flag p, CW w*
Output: *Unnecessary Handover Detection*
Init: $N\_pp, N\_cor \leftarrow 0$
*For each handover i:*
  $y\_p \leftarrow Pred\text{-}ToS(X\_i)$
  $short \leftarrow y\_p < ToS\_th$
  $osc \leftarrow |RSRP\text{-}Slope\_i[-1]| > rsrp\_th$ OR $|SNR\text{-}Slope\_i[-1]| > snr\_th$

  *is-pp ← short AND (osc OR y_p < osc_th)*
  *Adj is-pp with w[p_i]*
  *If p_i = 1 AND is-pp:*
   *N_cor, N_pp ← N_cor + 1, N_pp + 1*
*Return: N_pp, N_cor*

### 3.2.2 Avoidance of Unnecessary and Ping-Pong Handovers

If an event is detected as ping-pong or unnecessary, the avoidance logic evaluates the user trajectory and 3GPP signal safety thresholds to decide whether to suppress the handover as outlined in the avoidance algorithm.

**Algorithm 2: Avoidance**
Input: Seq X (RSRP, Bearing), ToS y_p, PP is-pp, THs θ_rsrp, θ_tos
Output: Handover Decision
Init: N_avd, N_maway ← 0
For each handover i:
 *maway ← |Bearing_i[-1] - Bearing_i[-2]| > 45°*
 *safe ← RSRP_i > θ_rsrp //3GPP*
 *unnec ← (y_p < θ_tos OR maway) AND safe*
 *If is-pp OR unnec:*
  *Avoid handover*
  *N_avd ← N_avd + 1*
 *Else:*
  *Exec handover //3GPP*
*Return: N_avd, N_maway*

### 3.3 Model Training and Optimization

GRU, LSTM, and Transformer architectures were trained on fixed-length sequences of serving/neighboring cell measurements, slope-derived indicators, and trajectory features to predict ToS and detect ping-pong events. Temporal dynamics were encoded as session-level elapsed time, while RSRP- and SNR-slopes captured signal variability. Features were Min–Max normalized to [0, 1]. Class imbalance, arising from the rarity of ping-pong events, was addressed through inverse-frequency weighting. Models were optimized with Adam, dropout, and early stopping; sequence length, hidden dimension, and learning rate were tuned via grid search. Evaluation include accuracy and F1-score for detection, reductions in ping-pong and unnecessary handovers with ToS gains for network utility, and training/inference time with model size for efficiency, enabling a rigorous cross-architecture comparison.

## 4. Results and Discussion

Table 1 summarizes the comparative performance of GRU, LSTM, and Transformer models across two feature sets: RSRP-only and all-feature inputs (RSRP, SNR, slopes, ToS, and bearing). The evaluation was conducted on five key metrics: ping-pong reduction, overall handover reduction, improvement in ToS, training time, and inference time.

### 4.1 Models Performance

The results highlight a clear advantage of using the full feature set over RSRP-only baselines. With RSRP-only inputs, GRU achieved the best ping-pong reduction (87.04%) but suffered from a limited precision–recall balance (F1 = 39.66). In this case, the model exhibited high precision but low recall, meaning that while most predicted unnecessary handovers were correct, many true cases were missed. In mobility management, sacrificing recall is tolerable, since a slightly higher number of missed ping-pongs is safer than wrongly blocking valid handovers that should occur. By contrast, when leveraging all features, GRU achieved a near-optimal 98.15% ping-pong reduction with an improved F1 of 45.10, representing the most robust trade-off between detection accuracy and false alarm mitigation. LSTM and Transformer models also benefited from feature enrichment, though to a lesser extent, indicating that recurrent structures, particularly GRU, are better suited to capturing the temporal dependencies inherent in handover sequences.

### 4.2 Handover Reduction and ToS Improvement

As shown in Figure 4, across models, integrating multi-dimensional features substantially improved handover reduction and ToS. GRU again dominated with a 46.25% reduction in unnecessary handovers and 32.53% ToS improvement, underscoring its ability to balance mobility robustness and service continuity. While LSTM achieved competitive handover reduction (40.31%) and ToS improvement (29.26%), its extended training and inference times make it less attractive for real-time deployment. The Transformer, although efficient in inference, achieved relatively lower reductions, suggesting its self-attention mechanism may be less sensitive to fine-grained mobility transitions compared to recurrent networks.

### 4.3 Computational Efficiency

Training time remains a critical bottleneck for LSTM Figure 5, which requires nearly 2× more training time than GRU under the all-feature setup (457.80 vs. 238.35). Inference latency is also unfavorable for

LSTM (2.66s) Figure 6, whereas GRU (0.91s) and Transformer (0.98s) maintain real-time feasibility. Importantly, GRU achieves this efficiency without sacrificing predictive robustness, positioning it as the most deployable model in latency-sensitive mobile networks

Table 1 Result Summary

|  | Model | Ping Pong Reduction | Ping Pong F1 | Handover Reduction | ToS Gain |
|---|---|---|---|---|---|
| RSRP based | GRU | 87.04 | 39.66 | 22.19 | 20.48 |
|  | LSTM | 79.63 | 43.65 | 28.75 | 26.20 |
|  | Transformer | 72.22 | 43.57 | 26.88 | 25.85 |
| All-features | GRU | 98.15 | 45.10 | 46.25 | 32.53 |
|  | LSTM | 98.15 | 42.23 | 40.31 | 29.26 |
|  | Transformer | 96.30 | 42.10 | 39.38 | 28.78 |

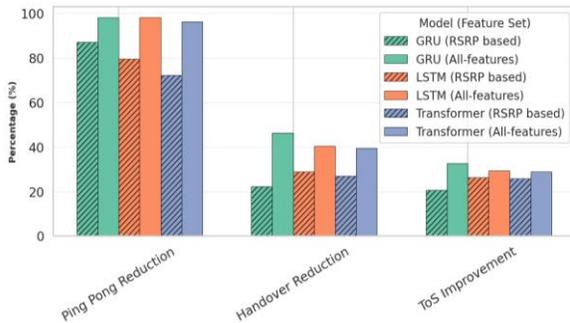

Figure 4 Handover Management Metrics Result

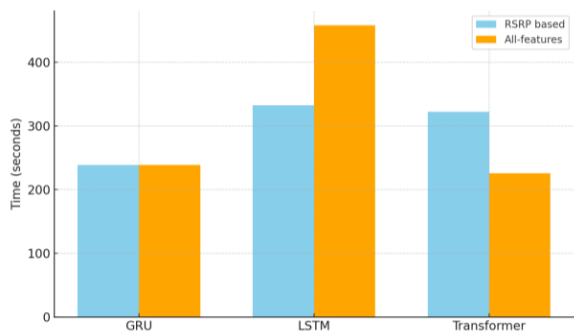

Figure 5 Training Time Comparison

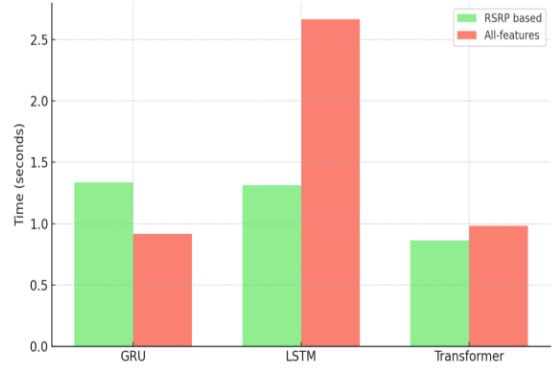

Figure 6 Inference Time Comparison

### 4.4 Implications for Mobility Management

The findings demonstrate that feature-rich sequential models, particularly GRU, can effectively mitigate unnecessary handovers by combining signal quality, trajectory, and temporal dynamics. This not only addresses ping-pong and short-stay issues but also enhances ToS a critical metric for QoE in dense cellular deployments. The improvements achieved (up to 98% ping-pong suppression and 46% handover reduction) substantially exceed those reported in recent deep learning-based mobility studies like [3], where performance gains are typically lower. This highlights the practical viability of integrating lightweight recurrent models into real-world mobility management entities for proactive decision-making.

While GRU shows strong adaptability, the performance gap between RSRP-only and all-feature models underscores the importance of holistic feature engineering. Future generalization across heterogeneous networks (e.g., 5G NR and beyond) may require expanding the feature space further (beam quality, mobility load, interference indicators) while retaining model simplicity. Additionally, although Transformers did not outperform GRUs in this study, their scalability and parallelism remain promising for large-scale training in operator-grade datasets.

### 5. Conclusion and Future Work

This study addressed the persistent challenge of unnecessary handovers and ping-pong effects in dense urban cellular networks by leveraging a large-scale, multi-operator drive-test dataset and formulating handover management as a sequence prediction problem. Through extensive evaluation of GRU, LSTM, and Transformer architectures under both RSRP-only and all-feature settings, the results demonstrate that feature-rich setup particularly GRU offer substantial performance gains. The GRU achieved up to 98% reduction in ping-pong events, 46% suppression of unnecessary handovers, and the lowest inference latency (0.91 s), establishing its suitability for real-time edge deployment. While LSTM provided a reasonable trade-off between complexity and accuracy, its higher computational burden limits its practicality, whereas the Transformer offered scalability advantages but fell short in mobility sensitivity compared to recurrent models.

A key insight from this study is that in handover management, precision must be prioritized over recall to preserve service continuity, with missed detections being less detrimental than false positives that could block valid handovers. Overall, the proposed framework significantly outperforms the conventional 3GPP A3 algorithm, yielding tangible improvements in mobility robustness, ToS, and user experience. The release of the dataset and code is expected to catalyze further research, enabling the design of adaptive, learning-driven mobility solutions for 5G and beyond.